\documentclass[11pt,twoside]{article}
\usepackage{asp2006}
\usepackage{lscape}
\usepackage{natbib}
\usepackage{color}

\usepackage{xspace}

\newcommand{\sn}{SN~Ia\xspace}
\newcommand{\sne}{SNe~Ia\xspace}

\def\edcomment#1{\iffalse\marginpar{\raggedright\sl#1\/}\else\relax\fi}
\reversemarginpar

\def\aj{AJ}%
\def\araa{ARA\&A}%
\def\apj{ApJ}%
\def\apjl{ApJ}%
\def\apjs{ApJS}%
\def\apss{Ap\&SS}%
\def\aap{A\&A}%
\def\nat{Nature}%
\begin{document}
\title{Modeling the Diversity of Type Ia Supernova Explosions}
\author{F. K. R{\"o}pke,  W. Hillebrandt,}
\affil{Max-Planck-Institut f{\"u}r Astrophysik,
  Karl-Schwarzschild-Str.~1, D-85741 Garching, Germany}
\author{D. Kasen and S. E. Woosley}
\affil{Department of Astronomy and Astrophysics, University of
  California, Santa Cruz, CA 95064, USA}

\begin{abstract}
Type Ia supernovae (SNe Ia) are a prime tool in observational
cosmology. A relation between their peak luminosities and the shapes of
their light curves allows to infer their intrinsic luminosities and to
use them as distance indicators. This relation has been established
empirically. However, a theoretical understanding is necessary in
order to get a handle on the systematics in SN Ia cosmology. Here, a
model reproducing the observed diversity of normal SNe Ia is
presented. The challenge in the numerical implementation arises from 
the vast range of scales involved in the physical
mechanism. Simulating the supernova on scales of the exploding white
dwarf requires specific models of the microphysics involved in the
thermonuclear combustion process. Such techniques are discussed and
results of simulations are presented.
\end{abstract}

\vspace{-0.5cm}
\section{Introduction}
\sne are extremely bright cosmic explosions with properties that are
more homogeneous than those of other astronomical
transients. Furthermore, a correlation between the width and the decline rate of
the $B$-band light curve (``width-luminosity relation'', WLR) allows
to calibrate them \citep{phillips1993a,phillips1999a} and makes them
the best distance indicators out to redshifts of about one. This
correlation, however, is established only empirically on a set of
nearby \sne. Consequently, a quantification of systematic errors
resulting from the calibration procedure is difficult to
achieve. Moreover, aiming at distance determinations to far events,
evolutionary effects could potentially obscure the 
measurements. It is clear that a sound understanding of the
physics of \sne is desirable in order to improve their precision as distance
indicators in observational cosmology. 

Besides adding to the ever growing cosmological \sn databases, observations in
the past decade allowed to take a close look at a number of nearby
events. It turned out that, although they form a remarkably
homogeneous class, individual \sne differ significantly in their
properties \citep[e.g.][]{benetti2005a, mazzali2007a}. Apart from variations within the ``normal''
\citep[as defined by][]{branch1993a} \sne, there are distinct sub-classes which differ
significantly from the bulk of events. A first step towards a physical understanding of the class of \sne and
to improve their quality as distance indicators is therefore to
identify the origin of this diversity. 

A model that reproduces a large range of observational characteristics
will be discussed in the following. Recent numerical simulations
suggest that it potentially can account for the ``normal'' \sne. This,
however, also implies that the distinct sub-classes have to be
explained in different physical scenarios.

\section{Astrophysical Model}

\sne are attributed to thermonuclear explosions of carbon-oxygen white dwarf (WD)
stars. In order to evolve a WD to a state where a thermonuclear
explosion can trigger, the supernova progenitor has to be a
binary system. Different scenarios of the progenitor evolution have been
suggested \citep[see, e.g.,][]{hillebrandt2000a} and may account for
different \sn subclasses. Here, we focus on the so-called
Chandrasekhar-mass explosion scenario, where the WD has accreted
matter from its companion so that its mass approaches the stability
limit -- the Chandrasekhar mass of $\sim$$1.4\, M_{\odot}$. Close
to this mass limit, the density in the core of the WD increases
dramatically so that carbon burning ignites. This, however, does not
yet trigger the actual explosion process because convective cooling still
moderates the burning. The resulting pre-explosion simmering phase
lasts for about a century \citep{woosley2004a} and is characterized by
highly turbulent convective motions. Gradually, the background temperature increases
so that finally one or many hotspots in the turbulent flow near the
WD's center undergo a
thermonuclear runaway. Out of these "ignition sparks", a flame starts to
propagate. It incinerates the WD material and leads to an explosion of
the star on time scales of $1$--$2 \, \mathrm{s}$. As discussed in
Sect.~\ref{sect:res}, the flame propagation is highly
sensitive to the initial conditions and the ignition
geometry strongly affects the strength of the overall
explosion process. Despite some recent progress \citep{
  hoeflich2002a, woosley2004a,
  wunsch2004a, kuhlen2006a, zingale2009a}, the physics
of the ignition process and the 
turbulent simmering phase is still extremely challenging and realistic
conditions are out of reach for numerical
implementations. Therefore, the number and spatiotemporal
distribution of ignition sparks remains uncertain.

After ignition near the center of the WD, the thermonuclear burning
front propagates outwards. Two distinct modes of propagation, a
subsonic deflagration and a supersonic detonation
\citep[see][]{landaulifshitz6eng} are consistent with the conservation
laws of hydrodynamics. The first attempts to modeling \sn explosions
\citep{arnett1969a,arnett1971a}, however, showed that a prompt
detonation converts 
the WD almost entirely into iron-group elements. This is in
contradiction with the observed spectra of these events which show strong
intermediate-mass element (IME) features. Such elements (like Si, Ca,
and S) are
synthesized in thermonuclear explosive burning at low densities and
therefore the WD material must expand prior to incineration. This can
only be achieved if the flame propagates subsonically, i.e.\ as a
deflagration.  Before reaching low densities, however,
the flame burns the high-density core material to iron-group
elements, predominantly $^{56}$Ni. In its radioactive decay, this
isotope releases Gamma-rays which are scattered down to optical
wavelengths in the ejecta and make the supernova bright.

A laminar deflagration is a very slow process \citep[see][for laminar
flame speeds]{timmes1992a}. In order to
burn sufficient amounts of material to explode the WD, the flame
propagation must accelerate significantly. And indeed, such an
acceleration is to be expected as the flame propagation from the WD's
center outwards produces a buoyancy-unstable stratification of light
and hot ashes under dense fuel in the gravitational field of the
star. The ensuing Rayleigh-Taylor instability leads to shear flows
between rising plumes of burning material and fuel downdrafts with
Reynolds numbers of $\sim$$10^{14}$. As a consequence of shear
instabilities, a turbulent eddy cascade establishes. Down to the Gibson
scale at which the laminar flame speed equals the eddy velocity, the
flame interacts with turbulent eddies of this cascade. For most parts
of the explosion, the Gibson scale lies orders of magnitude above the
flame thickness which is only of the order of millimeters to
centimeters. Therefore, turbulent eddies wrinkle and corrugate the
flame on large scales without affecting the microphysics of the
burning. In this flamelet regime of turbulent combustion, a sufficient
flame acceleration is achieved to explode the WD
\citep[e.g.][]{reinecke2002d, gamezo2003a, roepke2005b,
  roepke2007c}. Only when the fuel density falls below 
$\sim$$10^7\,\mathrm{g}\,\mathrm{cm}^{-2}$ (due to the WD expansion
and in the outer layers of the star), the Gibson scale becomes smaller than the
broadening flame structure. In this late stage of the explosion,
turbulent eddies stir the internal flame structure mixing fuel and
ash. \citet{niemeyer1997a} suggested that this may cause the
transition from an initial subsonic deflagration to a supersonic
detonation in the delayed detonation \sn scenario
\citep{khokhlov1991a}. Microphysical studies \citep{lisewski2000b,
  woosley2009a} indicate that entering the distributed burning regime
alone may not be sufficient for causing a deflagration-to-detonation
transition (DDT), but in addition turbulent velocities of $\sim$$1000
\, \mathrm{km}\,\mathrm{s}^{-1}$ at scales of about $10 \,
\mathrm{km}$ are required in this late burning regime. Analyzing
three-dimensional simulations of deflagrations in WDs,
\citet{roepke2007d} found that such high turbulent fluctuations do
occur in late stages of the explosion.

\section{Numerical Techniques}

The challenge in the numerical implementation of the above
astrophysical scenario of thermonuclear supernova explosions arises
from the multi-scale character of the problem
\citep[e.g.][]{roepke2008b, roepke2008d}. The physically
relevant range in spatial scales covers about 11 orders of
magnitude. In particular, the representation of a thin thermonuclear
flame on a grid comprising the entire WD and the flame/turbulence
interaction require special numerical approaches \citep[for a summary
see][]{roepke2009a}. 

In the implementation used here, 
the flame propagation is modeled based on the level-set technique
\citep{reinecke1999a}. It is associated with the zero-level set of a
signed distance function $G$ which is defined positive in the ashes
and negative in the fuel. The advancement of the flame is described by
a partial differential equation modifying this $G$-field in an
appropriate way \citep[see][]{reinecke1999a}. As an input, this requires the flame propagation
speed, which, in the flamelet regime, is set by the turbulent velocity
fluctuations on the scale of computational grid. These are derived
from a subgrid-scale turbulence model \citep{niemeyer1995b,schmidt2006c}.

Following \citet{golombek2005a} and \citet{roepke2007b}, the
detonation wave is propagated by a separate level-set function. This
allows to prevent the detonation from unphysically crossing ashes left
behind from the previous deflagration stage \citep{maier2006a}.

Another challenge arises from the expansion of the WD and the \sn
ejecta. In order to reliably compute synthetic observables from the
results of explosion simulations, these have to be followed to a state
of hydrodynamically relaxed homologous expansion. This requires to
simulate the explosion process for at least $10\, \mathrm{s}$. An
computational grid co-expanding with the ejecta allows to keep then on
the domain \citep{roepke2005c} and with moving nested grids the flame
front can be optimally resolved for a fixed number of grid cells \citep{roepke2006b}.

Another scale problem arises from the discrepancy between hydrodynamic
time scales and the time scales of some of the involved nuclear
reactions. Currently, this problem is bypassed in our implementation
by only following a very limited set of nuclear species in the
explosion simulation and reconstructing the details of the
nucleosynthesis in a post-processing step
\citep[e.g.][]{travaglio2004a, roepke2006b}. This is achieved on the
basis of passive tracer particles advected with the flow of the
explosion which record trajectories of temperature, energy, and
density. Both this tracer method and the expanding grid add Lagrangian
components to our Eulerian formulation of the problem.   

\section{Results}
\label{sect:res}

Based on the numerical techniques described above, thermonuclear
supernova explosion simulations in 
multiple spatial dimensions have been performed. While pure turbulent
deflagrations are able to explode the WD, the resulting event is
predicted to be on the faint end of the observed normal \sne \citep{roepke2007c}. 

The
delayed detonation scenario, in contrast, covers the range of normal to
bright \sne when the ignition of the deflagration is treated as a
stochastic process \citep{roepke2007b, mazzali2007a}. In cases of few
\citep{garcia2005a,livne2005a,roepke2006a} or asymmetrically distributed 
ignition sparks \citep[e.g.][]{calder2004a,roepke2007a,townsley2007a}, the deflagration is
weak. It burns comparatively little material and hence the production
of $^{56}$Ni and the energy release are low. Therefore, when the
detonation triggers, it finds abundant unburnt material at high
densities which it converts to additional $^{56}$Ni. The resulting
event is bright and energetic and the ejecta structure is set mainly
by the detonation stage. Somewhat counter-intuitively, a vigorous
ignition in many sparks distributed around the WD's center and a
subsequent strong deflagration leads to an overall faint
explosion. Here, the detonation finds an almost completely burnt core
and outer layers that have been diluted by expansion. It therefore
burns them predominantly to IMEs and does not significantly contribute
to the $^{56}$Ni production. Hence, the structure of the iron group
material at
the center of the ejecta is dominated by the large-scale buoyancy
instabilities from the deflagration. The outer IME layers, however,
are produced by the detonation and therefore smooth. 

A thorough
exploration of the deflagration ignition configuration as a
parameter of the delayed detonation scenario was performed in a set of
two-di\-mensional simulations \citep{kasen2009a}. Here,
spherical ignition kernels of radius $6 \, \mathrm{km}$ were placed
close to the center of the star -- in some models isotropically distributed
around it and in others in a solid angle with an opening of less than
360$^{\circ}$. In radius, a Gaussian distribution with a standard
deviation of $150 \, \mathrm{km}$ or $75 \, \mathrm{km}$ was
chosen. The number of ignition kernels ranged from 15 to 150
\citep[see the Supplementary Online Material of][]{kasen2009a}.
In addition to the
ignition spark distribution, the DDT criterion was varied. The
detonation was triggered in the distributed burning regime if the
ratio of the turbulent velocity fluctuation to the flame speed exceeded
a certain threshold. From the results of these simulations, synthetic
observables were computed by means of radiative transfer
calculations. These found good agreement of the models with
observations both in color light curves and in spectra and their
evolution. Furthermore, from these results the peak brightness and the
decline rate of the $B$-band light curve could be determined for the
individual models. These were found to follow the observational
relation of \citet{phillips1999a}.

\section{Conclusions}

Simulating the turbulent combustion in thermonuclear supernovae is a
numerically challenging task. Specific techniques are required to
correctly represent the thin flame and its interaction with turbulence
on a wide range of spatial scales on a computational grid that
comprises the full exploding WD. Here, a combination of a level-set
based flame representation and subgrid-scale turbulence modeling were
discussed as a solution to this problem. The implementation of this
approach allows to simulate thermonuclear supernova explosions. A pure
turbulent deflagration is found to be inconsistent with the
properties of ``normal'' \sne. These objects require a detonation stage
following an initial deflagration. The detonation wave is again
represented numerically with the level-set technique. Simulations of
the delayed detonation scenario lead to synthetic observables in good
agreement with the observations. If the ignition of the
deflagration is modeled as a stochastic process, the diversity of
luminosities found in normal \sne is reproduced. Moreover, the
correlation between peak luminosity and decline rate of the $B$-band
light curve is reproduced. This is the first time that
multidimensional hydrodynamical explosion models predict this
relation. Since it forms the basis of the calibration of \sne as
cosmological distance indicators, a theoretical understanding of the
correlation is required to improve the precision of \sn cosmology. Our
model provides the first step in this direction.

\acknowledgments{This research has been supported by the Emmy Noether Program
  of the German Research Foundation (DFG; RO~3676/1-1), the Cluster of
  Excellence EXC~153,
the Transregional Collaborative Research Center TRR~33,
the NASA Theory Program 
  (NNG05GG08G) and the DOE SciDAC Program (DE-FC02-06ER41438).}


\begin{thebibliography}{}
\expandafter\ifx\csname natexlab\endcsname\relax\def\natexlab#1{#1}\fi

\bibitem[{{Arnett}(1969)}]{arnett1969a}
{Arnett}, W.~D. 1969, \apss, 5, 180

\bibitem[{{Arnett} {et~al.}(1971){Arnett}, {Truran}, \&
  {Woosley}}]{arnett1971a}
{Arnett}, W.~D., {Truran}, J.~W., \& {Woosley}, S.~E. 1971, \apj, 165, 87

\bibitem[{{Benetti} {et~al.}(2005){Benetti}, {Cappellaro}, {Mazzali},
  {Turatto}, {Altavilla}, {Bufano}, {Elias-Rosa}, {Kotak}, {Pignata}, {Salvo},
  \& {Stanishev}}]{benetti2005a}
{Benetti}, S., {Cappellaro}, E., {Mazzali}, P.~A., {et~al.} 2005, \apj, 623,
  1011

\bibitem[{{Branch} {et~al.}(1993){Branch}, {Fisher}, \& {Nugent}}]{branch1993a}
{Branch}, D., {Fisher}, A., \& {Nugent}, P. 1993, \aj, 106, 2383

\bibitem[{{Calder} {et~al.}(2004){Calder}, {Plewa}, {Vladimirova}, {Lamb}, \&
  {Truran}}]{calder2004a}
{Calder}, A.~C., {Plewa}, T., {Vladimirova}, N., {Lamb}, D.~Q., \& {Truran},
  J.~W. 2004, arXiv:astro-ph/0405126

\bibitem[{{Gamezo} {et~al.}(2003){Gamezo}, {Khokhlov}, {Oran}, {Chtchelkanova},
  \& {Rosenberg}}]{gamezo2003a}
{Gamezo}, V.~N., {Khokhlov}, A.~M., {Oran}, E.~S., {Chtchelkanova}, A.~Y., \&
  {Rosenberg}, R.~O. 2003, Science, 299, 77

\bibitem[{{Garc{\'{\i}}a-Senz} \& {Bravo}(2005)}]{garcia2005a}
{Garc{\'{\i}}a-Senz}, D. \& {Bravo}, E. 2005, \aap, 430, 585

\bibitem[{{Golombek} \& {Niemeyer}(2005)}]{golombek2005a}
{Golombek}, I. \& {Niemeyer}, J.~C. 2005, \aap, 438, 611

\bibitem[{{Hillebrandt} \& {Niemeyer}(2000)}]{hillebrandt2000a}
{Hillebrandt}, W. \& {Niemeyer}, J.~C. 2000, \araa, 38, 191

\bibitem[{{H{\"o}flich} \& {Stein}(2002)}]{hoeflich2002a}
{H{\"o}flich}, P. \& {Stein}, J. 2002, \apj, 568, 779

\bibitem[{{Kasen} {et~al.}(2009){Kasen}, {R{\"o}pke}, \&
  {Woosley}}]{kasen2009a}
{Kasen}, D., {R{\"o}pke}, F.~K., \& {Woosley}, S.~E. 2009, \nat, 460, 869

\bibitem[{{Khokhlov}(1991)}]{khokhlov1991a}
{Khokhlov}, A.~M. 1991, \aap, 245, 114

\bibitem[{{Kuhlen} {et~al.}(2006){Kuhlen}, {Woosley}, \&
  {Glatzmaier}}]{kuhlen2006a}
{Kuhlen}, M., {Woosley}, S.~E., \& {Glatzmaier}, G.~A. 2006, \apj, 640, 407

\bibitem[{{Landau} \& {Lifshitz}(1959)}]{landaulifshitz6eng}
{Landau}, L.~D. \& {Lifshitz}, E.~M. 1959, Course of Theoretical Physics,
  Vol.~6, Fluid Mechanics (Oxford: Pergamon Press)

\bibitem[{{Lisewski} {et~al.}(2000){Lisewski}, {Hillebrandt}, \&
  {Woosley}}]{lisewski2000b}
{Lisewski}, A.~M., {Hillebrandt}, W., \& {Woosley}, S.~E. 2000, \apj, 538, 831

\bibitem[{{Livne} {et~al.}(2005){Livne}, {Asida}, \&
  {H{\"o}flich}}]{livne2005a}
{Livne}, E., {Asida}, S.~M., \& {H{\"o}flich}, P. 2005, \apj, 632, 443

\bibitem[{{Maier} \& {Niemeyer}(2006)}]{maier2006a}
{Maier}, A. \& {Niemeyer}, J.~C. 2006, \aap, 451, 207

\bibitem[{{Mazzali} {et~al.}(2007){Mazzali}, {R{\"o}pke}, {Benetti}, \&
  {Hillebrandt}}]{mazzali2007a}
{Mazzali}, P.~A., {R{\"o}pke}, F.~K., {Benetti}, S., \& {Hillebrandt}, W. 2007,
  Science, 315, 825

\bibitem[{{Niemeyer} \& {Hillebrandt}(1995)}]{niemeyer1995b}
{Niemeyer}, J.~C. \& {Hillebrandt}, W. 1995, \apj, 452, 769

\bibitem[{{Niemeyer} \& {Hillebrandt}(1997)}]{niemeyer1997a}
{Niemeyer}, J.~C. \& {Hillebrandt}, W. 1997, in NATO ASIC Proc., Vol. 486, NATO
  ASIC Proc. 486: Thermonuclear Supernovae, ed. P.~{Ruiz-Lapuente}, R.~{Canal},
  \& J.~{Isern} (Dordrecht: Kluwer Academic Publishers), 441--456

\bibitem[{{Phillips}(1993)}]{phillips1993a}
{Phillips}, M.~M. 1993, \apjl, 413, L105

\bibitem[{{Phillips} {et~al.}(1999){Phillips}, {Lira}, {Suntzeff}, {Schommer},
  {Hamuy}, \& {Maza}}]{phillips1999a}
{Phillips}, M.~M., {Lira}, P., {Suntzeff}, N.~B., {et~al.} 1999, \aj, 118, 1766

\bibitem[{{Reinecke} {et~al.}(2002){Reinecke}, {Hillebrandt}, \&
  {Niemeyer}}]{reinecke2002d}
{Reinecke}, M., {Hillebrandt}, W., \& {Niemeyer}, J.~C. 2002, \aap, 391, 1167

\bibitem[{{Reinecke} {et~al.}(1999){Reinecke}, {Hillebrandt}, {Niemeyer},
  {Klein}, \& {Gr{\"o}bl}}]{reinecke1999a}
{Reinecke}, M., {Hillebrandt}, W., {Niemeyer}, J.~C., {Klein}, R., \&
  {Gr{\"o}bl}, A. 1999, \aap, 347, 724

\bibitem[{{R{\"o}pke}(2005)}]{roepke2005c}
{R{\"o}pke}, F.~K. 2005, \aap, 432, 969

\bibitem[{{R{\"o}pke}(2007)}]{roepke2007d}
{R{\"o}pke}, F.~K. 2007, \apj, 668, 1103

\bibitem[{{R{\"o}pke}(2008)}]{roepke2008d}
{R{\"o}pke}, F.~K. 2008, in Modelling and Simulation in Science, ed.
  {V.~D.~Ges{\`u}, G.~Lo Bosco, \& M.~C.~Maccarone}, 74--82

\bibitem[{{R{\"o}pke} \& {Bruckschen}(2008)}]{roepke2008b}
{R{\"o}pke}, F.~K. \& {Bruckschen}, R. 2008, New Journal of Physics, 10, 125009

\bibitem[{{R{\"o}pke} {et~al.}(2006{\natexlab{a}}){R{\"o}pke}, {Gieseler},
  {Reinecke}, {Travaglio}, \& {Hillebrandt}}]{roepke2006b}
{R{\"o}pke}, F.~K., {Gieseler}, M., {Reinecke}, M., {Travaglio}, C., \&
  {Hillebrandt}, W. 2006{\natexlab{a}}, \aap, 453, 203

\bibitem[{{R{\"o}pke} \& {Hillebrandt}(2005)}]{roepke2005b}
{R{\"o}pke}, F.~K. \& {Hillebrandt}, W. 2005, \aap, 431, 635

\bibitem[{{R{\"o}pke} {et~al.}(2006{\natexlab{b}}){R{\"o}pke}, {Hillebrandt},
  {Niemeyer}, \& {Woosley}}]{roepke2006a}
{R{\"o}pke}, F.~K., {Hillebrandt}, W., {Niemeyer}, J.~C., \& {Woosley}, S.~E.
  2006{\natexlab{b}}, \aap, 448, 1

\bibitem[{{R{\"o}pke} {et~al.}(2007{\natexlab{a}}){R{\"o}pke}, {Hillebrandt},
  {Schmidt}, {Niemeyer}, {Blinnikov}, \& {Mazzali}}]{roepke2007c}
{R{\"o}pke}, F.~K., {Hillebrandt}, W., {Schmidt}, W., {et~al.}
  2007{\natexlab{a}}, \apj, 668, 1132

\bibitem[{{R{\"o}pke} \& {Niemeyer}(2007)}]{roepke2007b}
{R{\"o}pke}, F.~K. \& {Niemeyer}, J.~C. 2007, \aap, 464, 683

\bibitem[{{R{\"o}pke} \& {Schmidt}(2009)}]{roepke2009a}
{R{\"o}pke}, F.~K. \& {Schmidt}, W. 2009, in Interdisciplinary Aspects of
  Turbulence, ed. W.~{Hillebrandt} \& F.~{Kupka}, Lecture Notes in Physics
  (Berlin: Springer-Verlag), 255--289

\bibitem[{{R{\"o}pke} {et~al.}(2007{\natexlab{b}}){R{\"o}pke}, {Woosley}, \&
  {Hillebrandt}}]{roepke2007a}
{R{\"o}pke}, F.~K., {Woosley}, S.~E., \& {Hillebrandt}, W. 2007{\natexlab{b}},
  \apj, 660, 1344

\bibitem[{{Schmidt} {et~al.}(2006){Schmidt}, {Niemeyer}, {Hillebrandt}, \&
  {R{\"o}pke}}]{schmidt2006c}
{Schmidt}, W., {Niemeyer}, J.~C., {Hillebrandt}, W., \& {R{\"o}pke}, F.~K.
  2006, \aap, 450, 283

\bibitem[{{Timmes} \& {Woosley}(1992)}]{timmes1992a}
{Timmes}, F.~X. \& {Woosley}, S.~E. 1992, \apj, 396, 649

\bibitem[{{Townsley} {et~al.}(2007){Townsley}, {Calder}, {Asida}, {Seitenzahl},
  {Peng}, {Vladimirova}, {Lamb}, \& {Truran}}]{townsley2007a}
{Townsley}, D.~M., {Calder}, A.~C., {Asida}, S.~M., {et~al.} 2007, \apj, 668,
  1118

\bibitem[{{Travaglio} {et~al.}(2004){Travaglio}, {Hillebrandt}, {Reinecke}, \&
  {Thielemann}}]{travaglio2004a}
{Travaglio}, C., {Hillebrandt}, W., {Reinecke}, M., \& {Thielemann}, F.-K.
  2004, \aap, 425, 1029

\bibitem[{{Woosley} {et~al.}(2009){Woosley}, {Kerstein}, {Sankaran}, {Aspden},
  \& {R{\"o}pke}}]{woosley2009a}
{Woosley}, S.~E., {Kerstein}, A.~R., {Sankaran}, V., {Aspden}, A.~J., \&
  {R{\"o}pke}, F.~K. 2009, \apj, 704, 255

\bibitem[{{Woosley} {et~al.}(2004){Woosley}, {Wunsch}, \&
  {Kuhlen}}]{woosley2004a}
{Woosley}, S.~E., {Wunsch}, S., \& {Kuhlen}, M. 2004, \apj, 607, 921

\bibitem[{{Wunsch} \& {Woosley}(2004)}]{wunsch2004a}
{Wunsch}, S. \& {Woosley}, S.~E. 2004, \apj, 616, 1102

\bibitem[{{Zingale} {et~al.}(2009){Zingale}, {Almgren}, {Bell}, {Nonaka}, \&
  {Woosley}}]{zingale2009a}
{Zingale}, M., {Almgren}, A.~S., {Bell}, J.~B., {Nonaka}, A., \& {Woosley},
  S.~E. 2009, \apj, 704, 196

\end{thebibliography}

\end{document}